\documentstyle[prl,aps,epsfig,floats,psfrag]{revtex}
\begin{document}
\draft
\newcommand{\be}{\begin{equation}}
\newcommand{\ee}{\end{equation}}
\newcommand{\bea}{\begin{eqnarray}}
\newcommand{\eea}{\end{eqnarray}}
\def\lsim{\raise0.3ex\hbox{$\;<$\kern-0.75em\raise-1.1ex\hbox{$\sim\;$}}}
\def\gsim{\raise0.3ex\hbox{$\;>$\kern-0.75em\raise-1.1ex\hbox{$\sim\;$}}}
\def\Frac#1#2{\frac{\displaystyle{#1}}{\displaystyle{#2}}}
\def\no{\nonumber\\}
\def\slash#1{\ooalign{\hfil/\hfil\crcr$#1$}}
\def\ep{\eta^{\prime}}
\def\susy{\mbox{\tiny SUSY}}
\def\sm{\mbox{\tiny SM}}
\def\susy{\mbox{\tiny SUSY}}
\def\sm{\mbox{\tiny SM}}
\def\pmkz{\pi^-\overline{K}^0}
\def\pzkm{\pi^0{K}^-}
\def\ppkm{\pi^+{K}^-}
\def\pzkz{\pi^0\overline{K}^0}
\def\Kbar{\overline{K}}
\def\avebr{\overline{Br}}
\def\ddLL{(\delta^d_{23})_{LL}}
\def\ddLR{(\delta^d_{23})_{LR}}
\def\duLLone{(\delta^u_{31})_{LL}}
\def\duLLtwo{(\delta^u_{32})_{LL}}
\def\duLRone{(\delta^u_{31})_{LR}}
\def\duLRtwo{(\delta^u_{32})_{RL}}
\def\du#1#2{{\left(\delta^u_{#2}\right)_{#1}}}
\def\dd#1#2{{\left(\delta^d_{#2}\right)_{#1}}}
\newcommand{\W}{{\scriptscriptstyle W}}
\textheight      255mm  
\twocolumn[\hsize\textwidth\columnwidth\hsize\csname@twocolumnfalse\endcsname
\rightline{\small IPPP/04/42 \, \, DCPT/04/84 \, \, UCL-IPT-04-12}
\vskip0.5pc 
\title{Supersymmetric Contributions to $B\to K \pi$  Branching Ratio }
\author{S. Khalil$^{1,2}$, and E. Kou$^3$}
\address{$^1$~IPPP, University of Durham, South Rd., Durham
DH1 3LE, U.K.\\
$^2$~Ain Shams University, Faculty of Science, Cairo, 11566, Egypt.\\
$^3$~Institut de Physique Th\'{e}orique, Universit\'{e} Catholique de
Louvain, B1348, Belgium.\\}
\maketitle
\begin{abstract}
We analyze the supersymmetric contributions to the $B\to K\pi $ process. 
We show that the simultaneous contributions from the penguin diagrams with
chargino and gluino in the loop could lead to a possible solution to the
$B\to K \pi$ puzzle.
Our result indicates that including the stringent constraint from
the $b\to s \gamma$ branching ratio, the supersymmeteric
models with light right-handed top-squark and large mixing between second and 
third generation of up and down squarks is the most preferred by the
current experimental data.
\end{abstract}
\vspace*{-1.5cm}
]
\noindent 
Recent experimental measurements for the CP-averaged branching ratios of 
$B\to K \pi$ decays exhibit a possible discrepancy from the Standard Model 
(SM) prediction \cite{Babar,Belle}: 
\begin{eqnarray}
R_c &\equiv&2\left\{\frac{\avebr[B^+\to K^+\pi^0]+\avebr[B^-\to K^- \pi^0]}
{\avebr[B^+\to K^0 \pi^+]+ \avebr[B^- \to \bar{K}^0 \pi^-]} \right\}\nonumber\\
&=& (1.12 \pm 0.05)_{exp},\label{Rcresult}\\
R_n & \equiv & \frac{1}{2} \left\{\frac{\avebr[B^0 \to K^+ \pi^-] + 
\avebr(\bar{B}^0 \to K^- \pi^+)}{\avebr[B^0 \to K^0 \pi^0] + BR[\bar{B}^0 \to 
\bar{K}^0 \pi^0]} \right\}\nonumber\\
&=& (0.79 \pm 0.11)_{exp}.
\label{Rnresult}
\end{eqnarray}
As discussed in \cite{BF}, it is very difficult to have the situation 
$R_c>1$ and $R_n<1$ within the SM. 
While a confirmation with more accurate experimental data is necessary 
(an overestimate of $\pi^0$ can lead to a similar pattern of $R_c$ and $R_n$, 
see more in detail \cite{GR}), the current experimental values in Eqs. (\ref{Rcresult}) and (\ref{Rnresult}) 
does not seem to be possible even if we consider large hadronic uncertainties 
\cite{Yoshikawa,Kundu}. 
On the other hand, the fact that another combination of the branching ratios, 
\begin{eqnarray}
R&\equiv&\left\{\frac{\avebr[B^0\to K^+\pi^-] + \avebr[\bar{B}^0 \to K^- \pi^+]}
{\avebr[B^+\to K^0 \pi^+]+ \avebr[B^- \to \bar{K}^0 \pi^-]} \right\}\frac{\tau_B^+}
{\tau_{B^0}}\nonumber\\
&=& (0.89 \pm 0.07)_{exp}
\label{Rresult}
\end{eqnarray}
with $\frac{\tau_B^+}{\tau_{B^0}}=1.086\pm0.017$ \cite{PDG}, 
is almost consistent to the SM prediction leads us to the so-called 
largely enhanced electroweak penguin (EWP) mechanism \cite{BFRS} 
among various New Physics (NP) scenarios. Possible NP contributions 
to EWP have been studied (see e.g. in \cite{Trojan,BCLL,CFMMPS}) and also an impact on 
the future experiments of the $K$ decays are investigated \cite{BFRS,BSU}. 

In this letter, we analyze the supersymmetric contributions to the $B\to K\pi$ process in a 
model independent way using the mass insertion approximation. 
We will show that the $Z$ penguin diagrams with chargino in the loop contribute to the EWP 
significantly for a light right handed stop mass.  
Furthermore, the subdominant color suppressed EWP can be 
also enhanced by the the electromagnetic penguin ($O_{7\gamma}$) with chargino in the loop. 
The gluino contributions modify mainly the chromomagnetic penguins 
($O_{8g}$), i.e. the QCD penguins. As shown in \cite{KK1,CGHK}, this modified QCD penguin 
contribution would be the key to understand another hint of NP discovered in the 
$B$ factories; $S_{\phi K_S}\hspace*{-0.1cm}< \hspace*{-0.1cm}S_{J/\psi K_S}$.  
As we will show in this letter, the modified QCD penguin 
contributions would also help to deviate $R_c$ and $R_n$ in an indirect manner. 

In our computation, we apply the QCD factorization (QCDF) \cite{second,BN} which 
offers us an ability of estimating the hadronic matrix element of $O_{7\gamma}$. 
We extend the parameterization in \cite{second} 
by including the SUSY contributions. Then, Eq. (18) in \cite{second} can be rewritten as
\bea
A_{B^-\to\pmkz}&=&P\left[e^{i\theta_P}+\epsilon_ae^{i\phi_a}e^{-i\gamma}\right] \\
\sqrt{2}A_{B^-\to\pzkm}&=&P\big[e^{i\theta_P}+
\epsilon_ae^{i\phi_a}e^{-i\gamma}\nonumber\\
&-&\epsilon_{3/2}e^{i\phi}(e^{-i\gamma}-qe^{i\theta_q}e^{i\omega})\big] \\
A_{\overline{B}^0\to\ppkm}&=& P\big[e^{i\theta_P}+\epsilon_ae^{i\phi_a}e^{-i\gamma}\nonumber\\
&-&\epsilon_{T}e^{i\phi_T}(e^{-i\gamma}-q_Ce^{i\theta_{q_C}}e^{i\omega_C})\big] \\
-\sqrt{2}A_{\overline{B}^0\to\pzkz}&=&P\big[e^{i\theta_P}+\epsilon_ae^{i\phi_a}e^{-i\gamma} \nonumber\\
&+&\epsilon_{3/2}e^{i\phi}(e^{-i\gamma}-qe^{i\theta_q}e^{i\omega})\nonumber\\
&-&\epsilon_{T}e^{i\phi_T}(e^{-i\gamma}-q_Ce^{i\theta_{q_C}}e^{i\omega_C})\big] 
\eea
which satisfies the isospin relation:
\[
\sqrt{2}A_{\overline{B}^0\to\pzkz}=-A_{B^-\to\pmkz}+\sqrt{2}A_{B^-\to\pzkm}-A_{\overline{B}^0\to\ppkm}.
\]
The parameters $ \phi_a, \phi, \phi_T, \omega, \omega_C$ and $\theta_P, \theta_q, \theta_{q_C}$ 
are the CP conserving (strong) and the CP violating phase, respectively. 
The parameters are written as:
\begin{eqnarray*}
&&Pe^{i\delta_P}e^{i\theta_P}=\lambda_cA_{\pi\Kbar}[\alpha_4^c-\frac{1}{2}\alpha^c_{4,EW}+\beta_3^c+\beta^c_{3,EW}] 
\label{eq:P-def}\\
&&\epsilon_ae^{i\phi_a}=\Frac{\lambda_c\epsilon_{KM}}{P}\left[\beta_2+\alpha_4^u-\frac{1}{2}\alpha^u_{4,EW}+\beta_3^u+\beta^u_{3,EW}\right]\\
&&\epsilon_{3/2}e^{i\phi}=\Frac{-\lambda_c\epsilon_{KM}}{P}\left[ \alpha_1+R_{K\pi}\alpha_2+\frac{3}{2}(R_{K\pi}\alpha^u_{3,EW}+\alpha^u_{4,EW}) \right]\\
&&\epsilon_{T}e^{i\phi_T}=\Frac{-\lambda_c\epsilon_{KM}}{P}\left[ \alpha_1+\frac{3}{2}\alpha^u_{4,EW}-\frac{3}{2}\beta^u_{3,EW}-\beta_2) \right] \\
&&qe^{i\omega}e^{i\theta_q}\epsilon_{3/2}e^{i\phi}=\Frac{\lambda_c}{P}\left[\frac{3}{2}(R_{K\pi}\alpha^c_{3,EW}+\alpha^c_{4,EW})\right] \\
&&q_Ce^{i\omega_C}e^{i\theta_{q_C}}\epsilon_{T}e^{i\phi_T}=\Frac{\lambda_c}{P}
\left[\frac{3}{2}(\alpha^c_{4,EW}-\beta^c_{3,EW})\right]
\end{eqnarray*}
where $\lambda_u/\lambda_c\equiv\epsilon_{KM}e^{-i\gamma}$  and 
$R_{K\pi}=A_{\pi\Kbar}/A_{\Kbar \pi}\simeq 1.01 $.  
Comparing to \cite{second}, we have three extra CP violating phases, $\theta_P, \theta_q, \theta_{q_C}$ which may be induced 
by SUSY. 
$\alpha_{4 (3,EW; 4,EW)}^{u,c}$ contains the QCD  (EW) penguin contribution of both SM and SUSY. 
The SUSY contributions can be 
obtained by replacing SM Wilson coefficients in $\alpha_{4, (3,EW; 4,EW)}^{u,c}$ to the SUSY Wilson coefficients. 
$\alpha_1$ and $\alpha_2$ are the color allowed and 
color suppressed tree contributions, which do not contain SUSY contributions. $\beta_i^p$ represents the 
so-called weak annihilation contributions. We use the default values of all the input parameters in 
\cite{second} with $\rho_A=\rho_H=0$ 
in our numerical analysis. 
As a result, the SM contributions within QCDF are given by 
\begin{eqnarray*}
(Pe^{i\delta_P})_{\sm}&=& -0.0989 e^{-i 0.022}, (\epsilon_ae^{i\phi_a})_{\sm}=- 
0.0202 e^{i 0.25} \\
(\epsilon_{3/2}e^{i\phi})_{\sm}&=& 0.231 e^{-i 0.077},~~~~ (\epsilon_{T}
e^{i\phi_T})_{\sm}= 0.216 e^{i 0.00077 }\\
(qe^{i\omega})_{\sm}&=& \Frac{0.136 e^{-i 0.075}}{(\epsilon_{3/2}e^{i\phi})_{\sm}},~
(q_Ce^{i\omega_C})_{\sm}= \Frac{0.0143 e^{-i 0.88}}
{(\epsilon_Te^{i\phi_T})_{\sm}}.
\end{eqnarray*}
Notice that all the CP violating phases apart from $\gamma$ are zero in SM.  
These theoretical values lead to 
\be 
R_c=1.09(1.60),~ R_n=1.11(1.73),~ R=0.96(1.46) 
\ee
for $\gamma=\pi /3 (2\pi /3)$.  

We first introduce a parameterization in which 
the SUSY contribution manifests itself. 
By assuming the same strong phases for SM and SUSY, which is 
not a bad assumption in QCDF where the strong phases enter as  
higher-order correction, we can write 
\bea
P e^{i\theta_P}&=& P^{\sm}(1+ke^{i\theta^{\prime}_P}) \label{eq:19}\\
q e^{i\omega}e^{i\theta_q}&=& q^{\sm}e^{i\omega}(1+le^{i\theta^{\prime}_q}) \\
q_C e^{i\omega_C}e^{i\theta_{q_C}}&=& q_{C}^{\sm}e^{i\omega_C}(1+me^{i\theta^{\prime}_{q_C}}).  
\eea
where 
\bea
ke^{i\theta^{\prime}_P}&\equiv& \frac{(\alpha_4^c-\frac{1}{2}\alpha^c_{4,EW}+
\beta_3^c+\beta^c_{3,EW} )_{\susy}}{(\alpha_4^c-\frac{1}{2}\alpha^c_{4,EW}+
\beta_3^c+\beta^c_{3,EW} )_{\sm}}, \\ 
le^{i\theta^{\prime}_q}&\equiv& \frac{(R_{K\pi}\alpha^c_{3,EW}+\alpha^c_{4,EW})_{
\susy}}{(R_{K\pi}\alpha^c_{3,EW}+\alpha^c_{4,EW})_{\sm}}, \\ 
me^{i\theta^{\prime}_{q_C}}&\equiv& \frac{(\alpha^c_{4,EW}-\beta^c_{3,EW}) 
_{\susy}}{(\alpha^c_{4,EW}-\beta^c_{3,EW})_{\sm}}
\eea
The index SM (SUSY) means to keep only SM (SUSY) Wilson coefficients in 
 $\alpha_{i (,EW)}^p$ and $\beta_{i (,EW)}^p$.
$\epsilon_a$, $\epsilon_{3/2}$ and $\epsilon_T$ also include the QCD and EW penguin with 
the $u$ index. 
However, these are always suppressed by the factor 
$\epsilon_{KM}\simeq 0.020$ comparing to the SUSY contributions with the $c$ index. 
Therefore, neglecting this small contribution, the SUSY effects modify 
$\epsilon_{\{a,3/2,T\}}$  as 
\bea
(\epsilon_a e^{i\phi_a})&=&\frac{(\epsilon_a e^{i\phi_a})_{\sm}}{\left|1+k e^{i\theta^{\prime}_P}\right|}, \ \ 
(\epsilon_{3/2} e^{i\phi})=\frac{(\epsilon_{3/2} e^{i\phi})_{\sm}}{\left|1+k e^{i\theta^{\prime}_P}\right|}, \nonumber\\ 
(\epsilon_T e^{i\phi_T})&=&\frac{(\epsilon_T e^{i\phi_T})_{\sm}}{\left|1+k e^{i\theta^{\prime}_P}\right|} 
\eea

In order to have a general picture of the $B\to K\pi$ puzzle, an expanded formulae in terms of 
$\epsilon_T$, $\epsilon_{EW}\equiv q\times\epsilon_{3/2}$ and $\epsilon_{EW}^C\equiv q_C\times\epsilon_T$ is often useful. 
By assuming;  
i) the strong phases are negligible, i.e., $\phi_a, \phi, \omega, \omega_C$ are all zero, 
ii)the annihilation tree contribution is  negligible, i.e. $\epsilon_a\simeq 0$, 
iii)the color suppressed tree contribution is negligible,  i.e.  \hspace*{-0.1cm}$\epsilon_{3/2}e^{i\phi}\hspace*{-0.1cm}=\epsilon_Te^{i\phi_T}$ \hspace*{-0.25cm}, \hspace*{-0.2cm}
we can write $R_c$ and $R_n$ as 
\bea
R_c&\simeq&1+\epsilon_T^2-2\epsilon_T\cos (\gamma+\theta_P)+2\epsilon_{EW}\cos 
(\theta_P-\theta_q)  \no &-&2\epsilon_T\epsilon_{EW}\cos (\gamma +\theta_q) +
{\mathcal{O}}(\epsilon^2)\\
R_c-R_n&\simeq&2\epsilon_T\epsilon_{EW}\cos (\gamma +2\theta_P-\theta_q)\no
&-&2\epsilon_T\epsilon_{EW}^C\cos (\gamma +2\theta_P-\theta_{q_C})
+{\mathcal{O}}(\epsilon^2) \label{eq:RcRn} 
\eea

Now, let us find the configuration which leads to $R_c-R_n\gsim0.2$. 
From Eq. (\ref{eq:RcRn}), we can find that in general, the larger the 
values of $\epsilon_T$, $\epsilon_{EW}$ and $\epsilon_{EW}^C$ are, the larger 
the splitting between $R_c$ and $R_n$ we would acquire. 
Considering the SM value of $q_C\times \epsilon_T$, the second term in 
Eq. (\ref{eq:RcRn}) has only a tiny impact unless $m$ is of order 10.  
We can see that the phase combinations 
$\theta_P-\theta_q$ and $\theta_P+\gamma$ also play an important role. 
The possible solution to the $R_c-R_n$ puzzle by enhancing $\epsilon_{EW}$, which we 
have parameterized as $l$, has been intensively studied in the literature 
\cite{Yoshikawa,Kundu,BFRS}. 
As we will see in the following,  
$\epsilon_T$ can also be  enhanced when  
$ke^{i\theta_P^{\prime}}$ term contributes 
destructively against the SM and diminish $P$ (see Eq. (\ref{eq:19})). 
However, since $P$ is the dominant 
contribution to the $B\to K\pi$ process, the branching ratio is very sensitive 
to $ke^{i\theta_P^{\prime}}$. Therefore, we are allowed to vary 
$ke^{i\theta_P^{\prime}}$ only in the range of 
the theoretical uncertainty of QCDF, which gives about the right sizes of the 
$B\to K\pi$ branching ratios. Hence, we would be able to reduce $P$ at most by 
30 \%, which can be easily compensated by the error e.g. in the transition form 
factor $F^{B\to\pi, K}$. 

\begin{figure}[t]
\begin{center}\hspace*{-1cm}
\includegraphics[width=8cm]{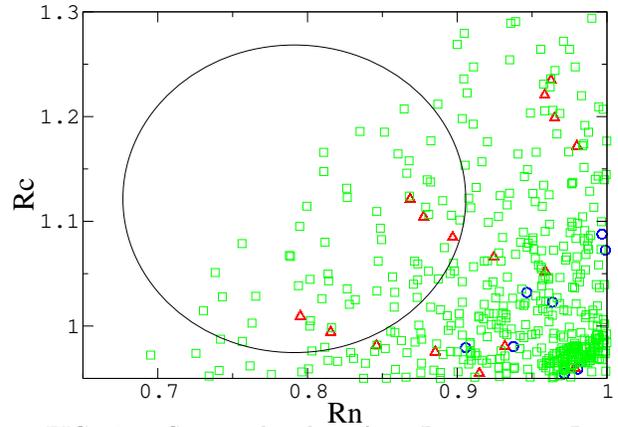}
\caption{Scattered plot for $R_n$ versus $R_n$ with 
$\{k=0, l=1, m=0\}$ (circle (blue)), $\{k=0, l=2, m=0\}$ (triangle (red)) and $\{k=0.3, l=1.5, m=0\}$ 
(square (green)) by varying $\gamma, \theta_p^{\prime}, \theta_{q}^{\prime}$. 
The ellipse indicates the experimental value.} 
\vspace{-0.6cm}
\end{center}
\end{figure}

Considering the tiny effect from the second term in Eq. (\ref{eq:RcRn}), in order to achieve 
$R_c-R_n\gsim 0.2$, we need $\epsilon_{T}\epsilon_{EW}$ larger than about 0.1 
or equivalently,  $\epsilon_{EW}$ larger than about 0.5 with $\epsilon_T^{\sm}$. 
In figure 1, we show scattered plots by varying $\gamma, \theta_p^{\prime}, 
\theta_{q}^{\prime}$ in a range of $\pi/3<\gamma <2\pi/3$ and $-\pi < 
\theta^{\prime}_{q_{(C)}} < \pi$ (interval of 0.2 for each phase). 
It should be noted that the figure is obtained by using the 
full formulae of $R_c$ and $R_n$ with the full QCDF result. 
The circles (blue), triangles (red) and squares (green) represent the result with 
$\{k=0, l=1, m=0\}$, $\{k=0, l=2, m=0\}$ and $\{k=0.3, l=1.5, m=0\}$, respectively.
 We can see that with $k=0$, we need $l\gsim 2$ to reproduce the experimental 
values while an inclusion of a small amount of $k$ lowers this bound significantly.

Now we investigate the SUSY contribution to 
the $B\to K \pi$ and the possible values of the parameters $k$ and $l$ to be reached. 
We work in the mass insertion approximation which is a technique to evaluate the SUSY 
contributions in a model independent manner. In this approximation, the extra 
FCNC as well as the new sources of CP violation enter in terms of the complex parameters 
$(\delta_{ij}^{q})_{AB}$ called mass insertion (MI) where $A,B$ are left (L) or right (R) indices, $i,j$ are generation 
indices and $q$ refers the up or down sector. While we will here try to constrain the MI 
from the experiments, it is also computed by given soft SUSY breaking models. 
As mentioned in the introduction, there are two kinds of new diagrams, one from the gluino loop 
and the other from the chargino loop. 
As can be found in \cite{GGMS}, 
the gluino contributions to the Wilson coefficients are proportional to $(\delta^d_{23})_{LL(RR)}$ 
apart from the magnetic terms, $C_{7\gamma}$ and $C_{8g}$, 
which additionally  receive contributions proportional to the chirality flipping $(\delta^d_{23})_{LR(RL)}$. 
For the moderate SUSY mass configurations we will consider in the following, 
we find that only the $(\delta^d_{23})_{LR(RL)}$ terms, 
which is enhanced by a factor $m_{\tilde{g}}/m_b$ comparing to the $(\delta^d_{23})_{LL(RR)}$ terms, 
play a significant role.  
As the chargino contributions, we consider the general SUSY models with light chargino mass 
as well as the case with light right handed stop. 
As given in \cite{CGHK}, the Wilson coefficients of the chargino diagrams  are proportional to 
the up sector MI and the leading term come from $\duLRtwo$ and $\duLLtwo$. 
The others $(\delta^u_{31})_{LL(RR)}$, $\duLRtwo$ and $(\delta^u_{32})_{RR}$ are 
Cabbibo suppressed by order $\lambda, \lambda^2, \lambda^3$, respectively where 
$\lambda\simeq 0.22$. The general formulae for the chargino Wilson coefficients are given as 
\begin{eqnarray}
F_{\chi}\!\simeq \!\du{LL}{32}\!R_F^{LL}\!+\!\du{RL}{32}\!Y_t R_F^{RL}\!\!,
\label{Fchapprox}
\end{eqnarray}
where the loop functions $R^{LL}_F$ and 
$R^{RL}_F$ can be found in Ref.\cite{CGHK}.
The index $F$ refers to $M^{\gamma}$ (electromagnetic-penguin ), 
 $M^{g}$(chromomagnetic-penguin), and $C$
($Z$-penguin ). It is important to note that the functions $R^{LL}_{M^{\gamma,g}}$ depend on 
the bottom Yukawa coupling $Y_b$ so that it can be enhanced for large $\tan\beta$. 
Also the function $R_C^{RL}$ largely increases when decreasing the mass of the 
right-stop. 
In the following, we consider only dominant three mass insertions, $\ddLR$, $\duLLtwo$ and 
$\duLRtwo$ discussed above, 
which have some enhancement factor and have a potential to lead to a large SUSY contributions 
to the $B\to K\pi$ process. 

Now let us show our numerical result.  As a default value of the SUSY parameters, we chose 
\bea
&m_{\tilde{g}}=500 \mbox{GeV}, \tilde{m}_{\tilde{q}}=500 \mbox{GeV}, &\no 
&m_{\tilde{t}_R}= 125 \mbox{GeV}, m_2=150 \mbox{GeV}, \mu= 250 \mbox{GeV} &\label{eq:mass}
\eea
First we present SUSY contributions to each 
$\alpha_{i,(EW)}^c$ from each MI.
The $\ddLR$ and $\duLLtwo$ terms which contribute to $C_{7\gamma}$ and $C_{8g}$ 
lead to   \vspace*{-0.2cm}
\[
\frac{\alpha_4^{c, \tilde{g}}}{\alpha_4^{c, \sm}}\simeq -36.7 \ddLR,~
\frac{\alpha_{4, EW}^{c, \tilde{g}}}{\alpha_{4, EW}^{c, \sm}}\simeq 27.7 \ddLR, 
\vspace*{-0.5cm}\] 
\[
\frac{\alpha_4^{c, \chi^+}}{\alpha_4^{c, \sm}}
\hspace*{-0.1cm}\simeq\hspace*{-0.1cm} -0.00110 \tan\beta \duLLtwo,~ 
\frac{\alpha_{4, EW}^{c, \chi^+}}{\alpha_{4, EW}^{c, \sm}}
\hspace*{-0.1cm}\simeq\hspace*{-0.1cm} 0.148 \tan\beta 
\duLLtwo, 
\vspace*{-0.2cm}\] 
respectively and the $\duLRtwo$ term which contributes to $Z$ penguin are obtained as 
\vspace*{-0.2cm}
\[
\frac{\alpha_{4, EW}^{c, Z}}{\alpha_{4, EW}^{c, \sm}}\simeq 1.68 \duLRtwo,~ 
\frac{\alpha_{3, EW}^{c, Z}}{\alpha_{3, EW}^{c, \sm}}\simeq 1.18 \duLRtwo .
\vspace*{-0.2cm}\]
Collecting all the SUSY contributions, our parameters $k, l,$ and $m$ are obtained as
\bea
ke^{i\theta_P}&=& -0.0019 \tan\beta \duLLtwo - 35.0 \ddLR +0.061\duLRtwo\label{eq:kSUSY}\no
le^{i\theta_q}&=& 0.0528 \tan\beta \duLLtwo-2.78\ddLR +1.11\duLRtwo \label{eq:lSUSY}\no
me^{i\theta_{q_C}}&=& 0.134 \tan\beta\duLLtwo + 26.4\ddLR +1.62 \duLRtwo \nonumber 
\eea
Note that we do not consider $(\delta^d_{23})_{RL}$ here 
but it is the same as $\ddLR$ with an opposite sign (see also \cite{KK2}).  

Let us first discuss the contributions from a single mass insertion 
$\duLLtwo, \ddLR$ or $\duLRtwo$ to $\{k, l, m\}$; 
keeping only one mass insertion and switching off the other two. 
Note that, as is well known, the absolute values of mass insertions $\duLLtwo$ and 
$\ddLR$ receive constraints from the $b\to s\gamma$ branching ratio which are 
 $|\tan\beta\times\duLLtwo|\leq 1$ and $|\ddLR|\leq 0.005$. 
 The mass insertion $\duLRtwo$ has only a constraint from its definition $|\duLRtwo|\leq1$. 
Firstly we discuss the $\duLRtwo$ term. Using $|\duLRtwo|=1$, 
the maximum value is found to be $\{k,l,m\}=\{0.061, 1.11, 1.62\}$. 
Thus, in this case where $k$ is almost negligible, 
we would need $l\simeq 2$ to explain the experimental data.  
We have a chance to enlarge the coefficients for $\duLRtwo$ by, for instance, 
increasing the averaged squark mass $\tilde{m}_{\tilde{q}}$. However, even if we choose  
$\tilde{m}_{\tilde{q}}=5$ TeV, we find that $l$ is increased only by 20 to 30 \%. 
Secondly, let us evaluate the $\ddLR$ and $\duLLtwo$ terms.  
Including the constraints from the $b\to s \gamma$ branching ratio, 
the maximum contributions from $\ddLR$ and $\duLLtwo$  are found to be  
$\{k, l, m\}=\{0.18, 0.014, 0.13\}$ and $\{0.0019, 0.053, 0.13\}$, which are far too small to explain the experimental data. 
The coefficients for $\ddLR$ depend on the overall factor $1/\tilde{m}_{\tilde{q}}$ and on also the variable of the loop function 
$x=m_{\tilde{g}}/\tilde{m}_{\tilde{q}}$ and we found that $m_{\tilde{g}}=\tilde{m}_{\tilde{q}}=250$ GeV can 
lead to  100 \%  increase. 
However, the value of $l$ is still too small to deviate $R_c-R_n$ significantly. As a whole, we found that it is extremely difficult to have  $R_c-R_n\gsim 0.2$ from a single mass insertion contribution. 

Let us try to combine two main contributions, $\ddLR$ and $\duLRtwo$ terms.  
Using the default SUSY masses in Eq. (\ref{eq:mass}) and including the $b\to s\gamma$ constraint to $|\ddLR|$, 
the maximum value is found to be $\{k, l, m\}=\{0.24, 1.12, 1.48\}$. 
The resulting $R_c-R_n$ are given as circles (blue) of Fig. 2 by varying $\arg\ddLR, \arg\duLRtwo, \gamma$. 
We can see that the experimental data are not reproduced very well. 
As discussed above, for a large value of the averaged squark masses,  $l$ increases while $k$ decreases. 
On the contrary, $k$ also depends on the ratio of gluino and squark masses. 
Hence we need to optimize these masses so as to increase $k$ and $l$ simultaneously. 
For instance, with 
$m_{\tilde{g}}\hspace*{-0.1cm}=\hspace*{-0.1cm}250$ GeV and $\tilde{m}_{\tilde{q}}\hspace*{-0.1cm}=\hspace*{-0.1cm}1$ TeV, we obtain   $\{k, l, m\}\hspace*{-0.1cm}=\hspace*{-0.1cm}\{0.30, 1.36, 1.90\}$ 
with which we find that quite a few points 
become well within the experimental bounds of $R_c$ and $R_n$ 
as shown in triangles (red) of Fig. 2. 

\begin{figure}[t]
\begin{center}
\includegraphics[width=8cm]{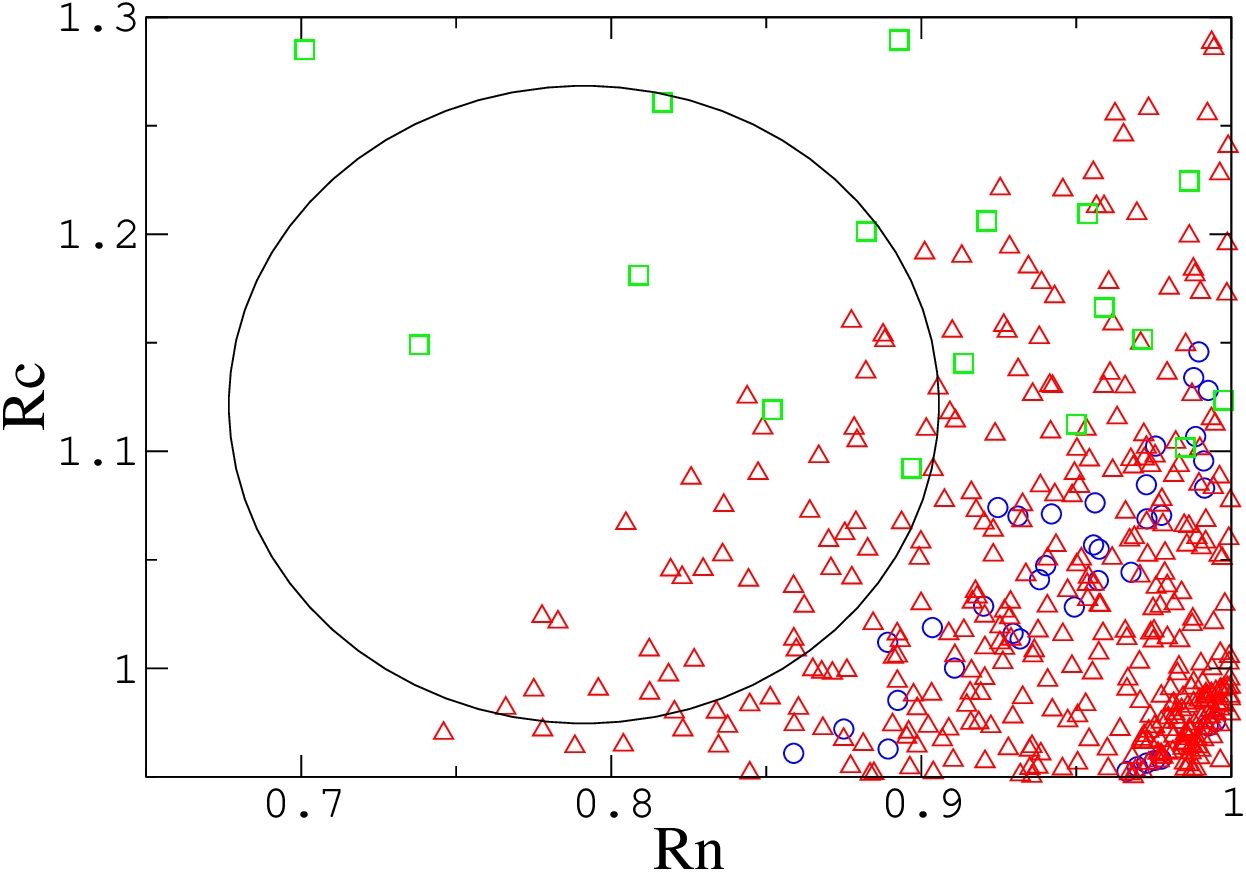}
\caption{Results of SUSY models with $|\duLRtwo|=1$ and $|\ddLR|=0.005$ 
(circle (blue) and  triangles (red); see the text for the other parameters) and  
$|\duLRtwo|=1$, $|\ddLR|=0.018$ and $|\tan\beta\times\duLLtwo|=6$ (square (green)). } 
\vspace{-0.6cm}
\end{center}
\end{figure}

Before concluding, let us mention about the relaxation of the $b\to s\gamma$  constraint 
as a possible solution when 
the experimental values of $R_c$ and $R_n$ remain around their current central values. 
The branching ratio of $b\to s\gamma$ depends not only on the absolute value of the 
mass insertion but also on the argument of it due to the overlap term of the SM and SUSY contributions. 
The constraints to the absolute value considered above are obtained independently from the 
phase. On the contrary, for a certain value of the argument,  
much larger  absolute value could be allowed. For example, 
$|\ddLR|=0.02$ with $\arg\ddLR \simeq \pm \pi$ and $|\tan\beta\times\duLLtwo|=6$ with $\arg\duLLtwo \simeq 0$ can also accommodate the experimental value of the $b\to s\gamma$ branching ratio. 
However, this does not help unfortunately since 
the constraint to $|\duLLtwo|$ is still too strict to deviate $R_c$ and $R_n$ 
significantly and also the phase condition for the $\ddLR$ term, $\arg\ddLR \simeq \pm \pi$, 
works in a direction of diminishing $R_c-R_n$. 
The situation is different once we consider the $\ddLR$ and $\duLLtwo$ terms simultaneously. 
When a condition $|\ddLR|={\mathcal{O}}(10^2)\times |\tan\beta\times\duLLtwo|$ is satisfied, 
the cancellation between the $\ddLR$ and $\duLLtwo$ terms occurs. 
Under this circumstance, 
we find that the same kind of relaxation for the absolute values can be expected but for much 
larger range of the phases comparing to the previous case. The squares (green) in Fig. 2 
show a scattered plot with  $|\duLRtwo|=1$, $|\ddLR|=0.018$ and $|\tan\beta\times\duLLtwo|=6$ by 
varying the phases in a rage of 
$\pi/3 < \gamma < 2\pi/3$ and $-\pi < \arg\duLRtwo, \arg\ddLR, \arg\duLLtwo < \pi$ 
{\it and excluding the points which  do not satisfy  the 
$b\to s\gamma$ branching ratio}. We  observe much larger $R_c\hspace*{-0.1cm}-\hspace*{-0.1cm}R_n$ 
in this scenario. 

In conclusion, we have examined the supersymmetric models as a solutions to the $B\to K\pi$ puzzle. 
We have shown that the $Z$ penguin diagram with chargino in the loop could be enhanced 
 for the small value of right handed stop mass and order one $\duLRtwo$.
We, however, found that this contribution itself is not large enough to solve the puzzle when choosing 
moderate values of the SUSY particle masses. 
We also found that $B\to K\pi $ receives a large gluino chromomagnetic penguin contributions 
($\ddLR$ term), which could explain another hint of new physics discovered in the B factories, 
$S_{\phi K_S}<S_{J/\psi K_S}$. 
While the $\ddLR$ term modifies only the QCD penguins and does not solve the $R_c-R_n$ 
puzzle directly, we found that it plays a complementary  role to the $\duLRtwo$ term and 
we can well explain the experimental values of $R_c$ and $R_n$ within the $b\to s\gamma$ constraint.  
The chargino electromagnetic penguin diagram  could also enhance EWP especially for 
large value of $\tan\beta$, however, we found that 
the stringent constraint to $|\tan\beta\times\duLLtwo|$ from the branching ratio of $b\to s\gamma$ 
prevents the $\duLLtwo$ term influencing  the values of $R_c$ and $R_n$. 
We examined a possible relaxation of the $b\to s\gamma$ constraint considering 
the phase configuration of the mass insertions. 
We found that the relaxation is quite possible especially, when a cancellation between 
the $\ddLR$ and $\duLLtwo$ term occur. 

As represented in the relaxation scenario, we found that 
the $R_c$ and $R_n$ data and furthermore the full spectrum of the $B\to K\pi$ branching ratios 
constrain the SUSY parameters for FCNC and CP violation very severely, which would 
provide a great opportunity of improving our knowledge of the SUSY breaking nature 
\cite{AKK}. 
 
\vspace{0.3cm}
\noindent {\bf Acknowledgments }\\ 
The authors acknowledge discussions with Emidio Gabrielli.  
The work by E.K. was supported by the Belgian
Federal Office for Scientific, Technical and Cultural Affairs through 
IAP P5/27. The work by S.K. was supported by PPARC.

\end{document}